\documentclass[prl,aps,amssymb,amsfonts,amsmath,twocolumn]{revtex4}
\pdfoutput=1

\usepackage{graphicx}
\usepackage{ucs} % for unicode characters

\usepackage{color} % for nice colors in the links
\definecolor{darkblue}{rgb}{0,0,0.5}
\definecolor{lila}{rgb}{0.3,0,0.3}
\definecolor{turq}{rgb}{0,0.1,0.4}
\definecolor{lightblue}{rgb}{0.7,0.7,0.9}

\usepackage{url} % that hyperlinks may be hyphenated correctly
\usepackage[pdftex,
  colorlinks=true,
  backref=page,
  linkcolor=darkblue, % usual links
  filecolor=red,
  citecolor=turq, % for bibliographic
  urlcolor=lila, % for Emails etc
  pdftitle={On-command control and imaging of single-molecule spectral dynamics using a nano-electrode},
  pdfauthor={Ilja Gerhardt, Gert Wrigge, Vahid Sandoghdar},
  pdfsubject={On-command control and imaging of single-molecule spectral dynamics using a nano-electrode},
  pdfkeywords={Single molecules, Stark effect, Spectral dynamics, NSOM, SNOM, Imaging, Near-field optics},
  pdfpagelabels=true, % damit die Nummerierung stimmt
  breaklinks=false,
  plainpages=false,
  backref=false,
  bookmarks,
  bookmarksnumbered=true]{hyperref}

\begin{document}

\title{Control and imaging of single-molecule spectral dynamics\\ using a nano-electrode}

\author{I. Gerhardt}\altaffiliation{present address: CQT, Centre for Quantum Technologies, 3 Science Drive 2, 117543 Singapore}
\author{G. Wrigge}
\author{V. Sandoghdar}
\affiliation{Laboratory of Physical Chemistry, ETH Zurich, CH-8093 Zurich, Switzerland}

\begin{abstract}
We study the influence of a scanning nano-electrode on fluorescence
excitation spectra of single terrylene molecules embedded in thin
\emph{p}-terphenyl films at cryogenic temperatures. We show that
applied voltages less than 10~V can result in reversible Stark
shifts of up to 100 times and linewidth increase greater than 10
times the natural linewidth. We discuss the potential of our
experimental scheme for direct imaging of individual two-level
systems (TLS) in the nanometer vicinity of single molecules.
\end{abstract}


\maketitle

Spectral dynamics (SD) is caused by the interaction of an emitter
with its surrounding matrix and is known to take place both in
crystalline and amorphous matrices. Depending on the time scale of
measurement, SD appears as discrete frequency jumps or broadening of
resonances. The nanoscopic origin of the emitter-matrix interaction
can vary in different cases and is in general poorly understood.
Nevertheless, the effects of the matrix can be formulated in a
phenomenological manner by considering the potential energy surface
that results from the large number of possible interactions between
the emitter and the constituents of the matrix. This energy
landscape can be treated as a reservoir of double-well
potentials~\cite{Heuer:1993} with their lowest energy levels
referred to as two-level systems (TLS)~\cite{Phillips:1972p351,
anderson_anomalous_1972,Geva:97}. Tunneling between the potential
wells gives rise to SD which can be mediated via mechanical,
magnetic, electric, or acoustic interactions. In what follows, we
restrict ourselves to the influence of external electric fields on
the SD of zero-phonon lines (ZPL) of organic dye molecules at
cryogenic temperatures.

Although in an amorphous matrix the emitter is expected to interact
with a sea of TLSs, it is possible that one or only few TLSs
dominate in a crystalline environment, leading to discrete jumps.
One such an example was discovered for terrylene molecules embedded
in \emph{p}-terphenyl (pT), in which guest molecules in certain
sites undergo well-defined spectral jumps under laser
illumination~\cite{kulzer_nature_1997,bordat_jchemphys_2002}.
Another case of light-induced reversible spectral jumps was
demonstrated for terrylene in naphthalene~\cite{bach02} by exciting
triplet states of the matrix molecules. Nanomechanical interactions
have also been shown to cause SD in a reversible fashion in recent
near-field spectroscopy experiments~\cite{Gerhardt:07b}. An elegant
avenue for external control of TLSs, which is of central concern in
this article, is via application of electric fields~\cite{Maier:95}.
Segura \emph{et al.} observed the narrowing of the ZPLs in two
individual pentacene molecules embedded in pT when they applied a
voltage to a micro-electrode put in contact with the
sample~\cite{segura_cpl_2001}. A related experiment was performed on
terrylene in \emph{n}-hexadecane by Bauer \emph{et al.}~\cite{bauer_jchemphys_2003}, who found many cases of double
resonances, which could be split further and experienced changes in
their peak intensities under electric fields. Both of the
above-mentioned experiments estimated large dipole moments up to 8
Debye for a TLS.

\begin{figure}[b!]
\centering
\includegraphics[width=70mm]{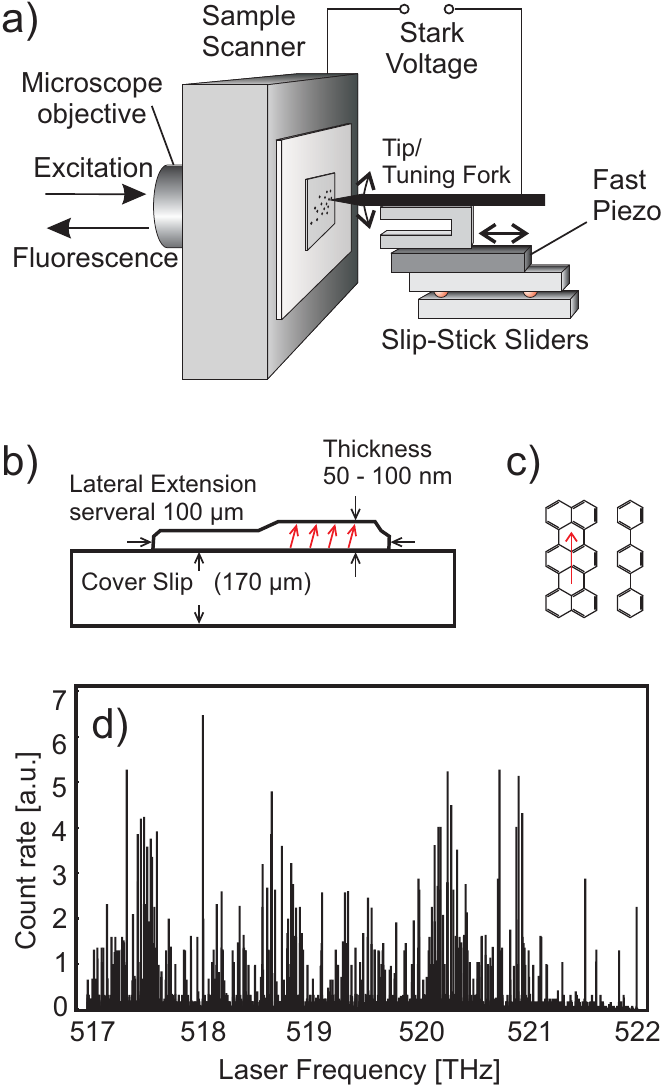}
\caption{(a) Schematics of the combined cryogenic near-field and
confocal microscope. In the current work, the microscope objective
was used both for illumination and collection of single-molecule
fluorescence. (b) Geometry of the spin-casted \emph{p}-terphenyl
thin film on a glass cover slide. (c) The structures of the dye
molecule terrylene and the host matrix \emph{p}-terphenyl. (d) The
inhomogeneously distributed excitation spectrum of the sample.
\label{fig:setup}}
\end{figure}

Exploration and control of the interaction between single emitters
and their nanoscopic environments provide very valuable insight into
various optical phenomena in the condensed phase, and they might
open doors for engineering new quantum systems. In this article, we
present experiments that employ metal-coated glass fiber tips as
scanning nano-electrodes to apply local electric fields on thin pT
samples doped with terrylene molecules. We report large
modifications of the ZPL linewidths and discuss future directions
for imaging single TLSs and their interaction with neighboring
single molecules.

\begin{figure}[t!]
\centering
\includegraphics[width=70mm]{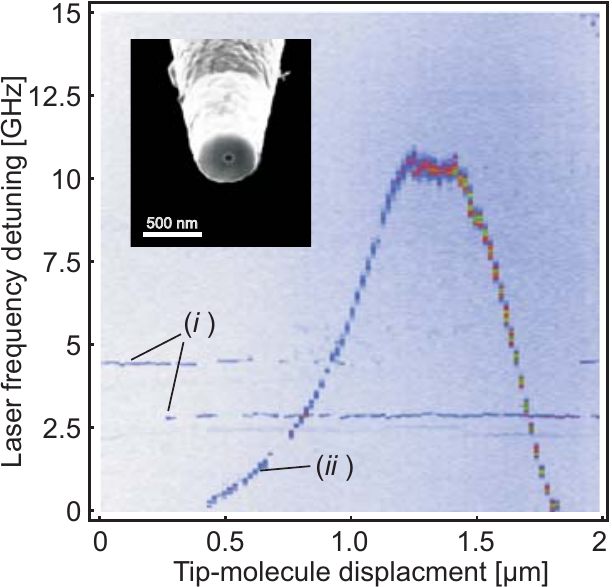}
\caption{Fluorescence excitation spectra of two molecules as a
function of tip position. Molecule (i) is far from the tip and
experiences spectral jumps. Molecule (ii) is strongly Stark-shifted
as the tip is scanned across it in a constant height. The inset
shows an electron micrograph of a FIB-milled metalized fiber tip.
\label{fig:lateral-scan}}
\end{figure}

In order to achieve a high spatial resolution in manipulating the
TLS, one requires large electric field gradients, ideally changing
the observed spectrum of the emitter by more than one linewidth over
a length scale of about one nanometer. As displayed in
Fig.~\ref{fig:setup}a, the machinery of scanning near-field optical
microscopy (SNOM) provides a convenient way for realizing this
scheme. To produce a well-defined SNOM probe, we pulled glass fiber
tip and coated them with aluminum and milled at the end by a focused
ion beam~\cite{veerman1998}. An electron micrograph of such a probe
is shown as an inset in Fig.~\ref{fig:lateral-scan}. Home-built
slip-stick sliders allowed us to position the tip at a desired
location with sub-micrometer accuracy. The probe-sample distance was
controlled via the shear-force interaction using commercially
available quartz tuning forks with a resonance frequency of
32~kHz~\cite{Karrai-95APL}. In this scheme, lateral (sub-)nanometer
oscillations of the tip are measured by a lock-in amplifier that
monitors the current created in the tuning fork. At very small
separations of a few tens of nanometers from the sample, the tip
oscillation is damped owing to the so-called shear-force effect.
This provides an error signal that is used to control the tip
position and to prevent if from crashing into the sample. A
piezo-electric scanner was used to scan the sample more than six
micrometers with nanometer accuracy in all directions. Further
details of our cryogenic confocal-SNOM setup can be found in
Refs.~\cite{Hettich:02,Gerhardt:07b}.

In our current work, we used the SNOM tip only as a nano-electrode
for applying a local electric field, while the cryostat body was
grounded and used as the counter-electrode (see
Fig.~\ref{fig:setup}a). Furthermore, we adjusted the tip-sample
separation at about 100~nm to rule out any mechanical
contact~\cite{Gerhardt:07b} or near-field effects related to the van
der Waals interaction~\cite{chance1978,bian_prl_1995}. Optical
measurements were not performed through the tip, as is usually done
in SNOM; instead, we used a microscope objective on the other side
of the sample to both excite the molecules and collect their
fluorescence in the far field. To record fluorescence excitation
spectra of a single terrylene molecule, we used a narrow-band dye
laser (Coherent Autoscan, 899-21, $\lambda \approx $ 577~nm) to
excite it selectively and detected its Stokes-shifted fluorescence
on an avalanche photodiode through an optical long-pass filter
(Omega-Optical, Kaiser notch-filter)~\cite{Orrit:90}.

Electric fields at the end of a sharp electrode drop rapidly at
distances larger than the tip
diameter~\cite{Jackson-book,Hettich:02}. Therefore, to ensure a
strong spatial gradient, it is important to place the molecule of
interest as close to the tip as possible. We, thus, performed our
experiments on thin pT
films~\cite{sandoghdar_sm_2001,pfab_chemphsylett_2004}. Briefly, a
pira\~na-cleaned cover slip was spin coated with a solution of pT in
toluene (3~mg/ml) doped with terrylene molecules at a concentration
of about $10^{-6}$. This resulted in typical film thickness of about
20-30~nm. However, cryogenic fluorescence excitation spectra
revealed spectral jumps and typical linewidths of several hundered
MHz, which are many times larger than the natural linewidth of 40
MHz for terrylene in pT. To minimize the residual SD, we annealed
the samples at 60\textdegree C for 60~min in air. As a result, pT
films were split to islands with thicknesses increased to 50-100~nm
(see Fig.~\ref{fig:setup}b). Contrary to what is known from bulk
samples~\cite{kummer_jchemphys_1997}, Fig.~\ref{fig:setup}d shows
that the inhomogeneous distribution of the ZPLs did not show
well-defined spectral sites in these islands. We mention in passing
that we observed a bunching of molecules near crystal island edges
in wide-field images. To avoid possible complications related to
edges, we performed our measurements on molecules several
micrometers from the rims. Furthermore, we occasionally witnessed
spontaneous cracks in the films even days after introduction of the
sample into the cryostat, pointing to internal strain in parts of
the sample. Nevertheless, the annealing procedure allowed us to
re-gain lifetime-limited linewidths of 40-50~MHz in some single
molecule spectra.

Figure~\ref{fig:lateral-scan} displays spectra recorded from two
molecules as a grounded tip was scanned laterally along a line and
at a constant height of 100~nm. Molecule (\emph{i}) is not directly
affected by the tip because it is not close enough to it. However,
we clearly see discrete and reversibly recurring spectral jumps
between two well-defined frequencies (at about 2.7 and 4.5~GHz on
the vertical axis). In case of molecule (\emph{ii}), the ZPL is
shifted by more than 10~GHz as the tip crosses the molecule. A
plateau of 300~nm clearly mirrors the tip shape. We also notice that
the largest frequency shift is accompanied by the broadening of the
resonance. It turns out that although the SNOM tip was grounded, the
large frequency shift encountered in Fig.~\ref{fig:lateral-scan} can
be explained by a Stark shift.

The Stark shift $W(\mathbf{E})$ produced by an external electric
field $\mathbf{E}$ can be expressed as
\begin{eqnarray}
W(\mathbf{E}) &=& W(0) - (\vec{\mu}_{\rm e} - \vec{\mu}_{\rm g}) \cdot (\mathbf{E+E_0})\nonumber \\&&- \frac{1}{2}(\alpha (\mathbf{E+E_0}) \cdot(\mathbf{E+E_0})) + \ldots
\end{eqnarray}

\noindent where $(\vec{\mu}_{\rm e} - \vec{\mu}_{\rm g})$ is the
difference between the electric dipole moments of the excited and
ground states, $\alpha$ is the difference in the polarizabilities of
the two states, and $\mathbf{E_0}$ represents a residual electric
field in the system~\cite{basche_book}. For a single terrylene
molecule in a centro-symmetric environment with no disorder
$\vec{\mu}=0$, so that $W(\mathbf{E})$ only shows a quadratic
dependence. In a host system with broken symmetry, however, the
linear effect can dominate. Furthermore, if there are any residual
electric fields in the sample, for example due to external charges,
the second term of Eq.~(1) can give rise to a contribution that is
linear in $\mathbf{E}$.

\begin{figure}[t!]
\centering
\includegraphics[width=70mm]{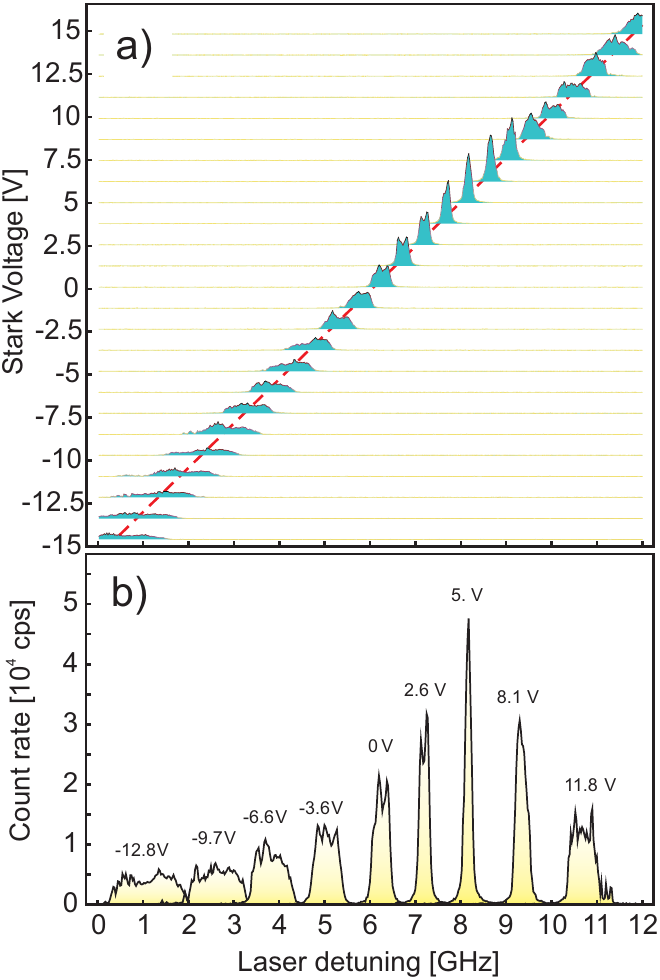}
\caption{(a) Linear Stark effect of a single molecule in the sample.
The area of the resonance line remains almost the same during the
voltage scan. At an applied voltage of 5~V the linewidth is minimal.
(b) A selection of line shapes at different
voltages.\label{fig:stark}}
\end{figure}

In Fig.~\ref{fig:stark}a, we varied the applied voltage to a
stationary nano-electrode placed above the molecule and recorded
fluorescence excitation spectra. Each frequency scan took 14.4~s at
a resolution of 25~MHz and integration time of 30~ms per frequency
pixel. We find a linear blue Stark shift for increasing positive
voltages. However, as reported in Ref.~\cite{bauer:10278}, we have
also observed the opposite effect for different molecules under the
same tip. A similar spectral behavior has been observed for the
XY-site of terrylene in pT~\cite{kulzer_jphyschema_1999}, but as
mentioned earlier, our samples do not show any evidence of
well-defined sites (see Fig.~\ref{fig:setup}d). A rough estimate
yields effective field strengths up to $\pm$1~MV/cm applied to the
molecule, which is about two orders of magnitude stronger than in
earlier single molecule
experiments~\cite{segura_cpl_2001,bauer_jchemphys_2003,bauer:10278,Bauer2004}.

Interestingly, the observed linear Stark effect in
Fig.~\ref{fig:stark}a is accompanied by a change in the spectral
line shape, while the area below the resonance remains constant to
within approximately 20\% over the whole voltage range of -15 to
+15~V. Figure~\ref{fig:stark}b depicts a few examples for a closer
scrutiny. We find that the resonance line becomes a narrow
Lorentzian shape with a full width at half-maximum of 130~MHz at an
applied voltage of about 5~V. The data in Figs.~\ref{fig:stark}a
and~\ref{fig:lateral-scan} reveal that for this molecule, the
unperturbed ZPL of 66~MHz is broadened if a grounded tip is placed
above it. It is possible that our nano-electrodes are charged in the
fabrication process, which involves focused-ion-beam milling and
electron microscopy. An oxide layer formed on the aluminium-coated
tip during handling in air might embed these charges and protect
them even when the tip shaft is grounded. A nonzero residual
electric field might also exist in the pT sample. In any event,
Fig.~\ref{fig:stark} shows that application of 5~V minimizes the
effect of $\mathbf{E_0}$. At the same time, both positive and negative
fields of larger magnitudes broaden the line to well beyond its
value at $E=0$.

Clearly, a simple Stark effect is not expected to affect the
resonance line shape. We also ruled out any systematic effect of tip
oscillation and possible modulation of Stark shift by verifying that
the spectra in Fig.~\ref{fig:stark} did not change when we switched
off the 32~kHz dither of the tuning fork. On the other hand, the
strong response of the spectra to the applied voltages let us
attribute the observed broadenings to the electric field-induced SD
as reported in
Refs.~\cite{Maier:95,segura_cpl_2001,bauer_jchemphys_2003}. In this
frame, electric polarization of a TLS in the vicinity of the
molecule modifies the relative depths of its two potential wells.
The new energy landscape determines whether the tunneling process is
accelerated or slowed down in the presence of the tip. Broadenings
greater than 1.5~GHz at large applied voltages indicate the
activation of many TLSs and fast tunneling rates, contrary to the
cases encountered by Bauer et al.~\cite{bauer_jchemphys_2003}, where
single TLSs could be identified.

\begin{figure}[t!]
\centering
\includegraphics[width=70mm]{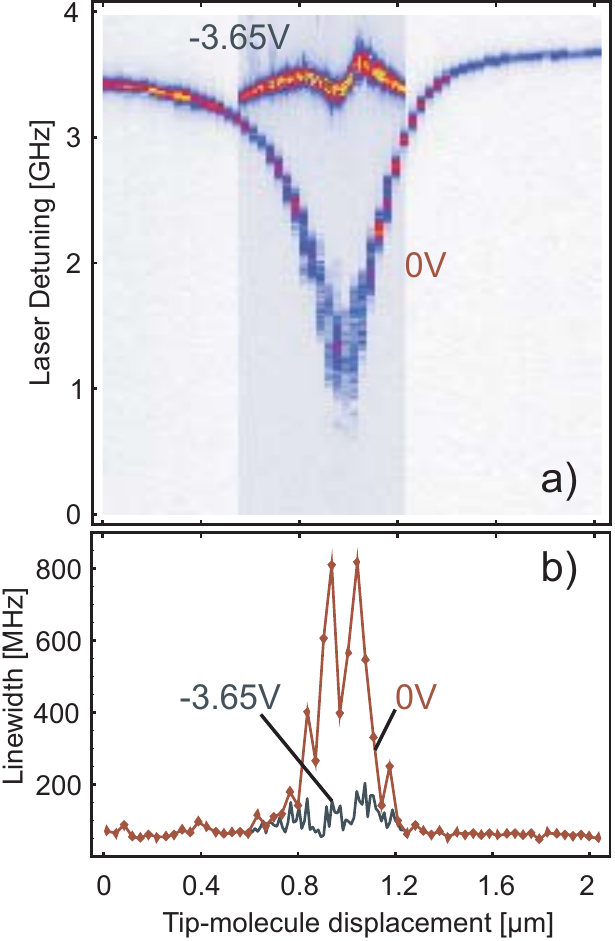}
\caption{a) Fluorescence excitation spectra recorded as a function
of the lateral tip position for voltages 0~V and -3.65~V applied to
it. b) The linewidths (full width at half-maximum) of spectra in a).
\label{fig:compensation}}
\end{figure}

Having discussed the mechanism of electric field-induced
broadenings, we turn our attention to Fig.~\ref{fig:compensation}a,
where spectra recorded on a different single molecule and with
another tip are presented. This time, we observed broadenings as
large as 800~MHz accompanying a shift of 1.8~GHz over a lateral tip
displacement of about 150~nm. However, as the upper curve of
Fig.~\ref{fig:compensation}a plots, both the shift and broadening of
the spectra were strongly reduced when a voltage of -3.65~V was
applied to the tip. In particular, we see that the originally
unperturbed linewidth of 50~MHz can be recovered, demonstrating a
reversible on-command manipulation of the TLS-molecule interaction.
The dip in the upper curve of Fig.~\ref{fig:compensation}a is most
likely caused by the opening in the metal coating of the SNOM tip
(see inset in Fig.~\ref{fig:setup}). As mentioned earlier,
near-field surface effects related to the van der Waals
interaction~\cite{chance1978,bian_prl_1995} are not expected to play
a significant role in our experiments because of a relatively large
molecule-tip separations of about 100~nm.

In a previous study, we have shown that the position-dependent Stark
shift produced by a scanning micro-electrode can be used to localize
single molecules with nanometer accuracy~\cite{Hettich:02}. We now
propose that the position-dependent broadening observed in
Figs.~\ref{fig:lateral-scan} and \ref{fig:compensation} can be
exploited to localize a TLS. The centers of the linewidth profiles
in these figures cannot be determined with good accuracy because of
a relatively large tip plateau and its aperture (see inset in
Fig.~\ref{fig:lateral-scan}). However, the large field gradient
created by the edge of the SNOM tip gives rise to the sharp slope of
22~MHz/nm, or equivalently, one natural linewidth per 2-3~nm. Use of
sharper tips and higher electric field gradients should provide
substantially higher position sensitivity, allowing one to resolve
individual TLSs in the vicinity of a molecule.

In conclusion, we have demonstrated \emph{in-situ} manipulation and
control of the phononic interaction of a single molecule with its
nano-environment. We have applied a scanning nano-electrode to show
that local application of static electric fields as small as 0-10~V
can not only induce large Stark shifts of several GHz in the optical
transition of an emitter, but also change its spectral linewidth.
Extension of our experimental scheme to sharper electrodes than
those used in this work will make it possible to image both a single
molecule and the TLSs in its immediate neighborhood simultaneously,
and thus, identify the origins of spectral dynamics at the molecular
level.

\vspace{3mm}

% Acknowledgement
We thank Joy Tharian and Phillipe Gasser at the Swiss Federal
Institute for Material Research (EMPA) for their help in producing
optical near-field tips and Robert Pfab and Enrico Klotzsch for
their contributions to the early stage of sample preparation. This
work was financed by the Schweizerische Nationalfond (SNF).

\end{document}